\documentstyle[preprint,aps,psfig]{revtex}	

\begin{document}
\draft
%
%
%
\title{The energy dependence of flow in Ni induced collisions\\
from 400 to 1970A $MeV$}
\renewcommand{\thefootnote}{\thefootnote}{\fnsymbol{footnote}}
\author{
 J.~Chance$^{(1)}$, S.~Albergo$^{(2)}$, F.~Bieser$^{(4)}$,
F.~P.~Brady$^{(1)}$,
Z.~Caccia$^{(2)}$, D.~Cebra$^{(1,4)}$, A.~D.~Chacon$^{(6)}$,
Y.~Choi$^{(5)}$\footnotemark, S.~Costa$^{(2)}$, J.~B.~Elliott$^{(5)}$,
M.~Gilkes$^{(3)}$\footnotemark, J.~A.~Hauger$^{(5)}$, A.~S.~Hirsch$^{(5)}$,
E.~L.~Hjort$^{(5)}$, A.~Insolia$^{(2)}$, M.~Justice$^{(3)}$,
D.~Keane$^{(3)}$,
J.~C.~Kintner$^{(1)}$\footnotemark, M.~A.~Lisa$^{(4)}$,  H.~S.~Matis$^{(4)}$,
M.~McMahan$^{(4)}$, C.~McParland$^{(4)}$, D.~L.~Olson$^{(4)}$,
M.~D.~Partlan$^{(1)}$\footnotemark, N.~T.~Porile$^{(5)}$, R.~Potenza$^{(2)}$,
G.~Rai$^{(4)}$, J.~Rasmussen$^{(4)}$,  H.~G.~Ritter$^{(4)}$,
J.~Romanski$^{(2)}$\footnotemark, J.~L.~Romero$^{(1)}$,  G.~V.~Russo$^{(2)}$,
R.~Scharenberg$^{(5)}$, A.~Scott$^{(3)}$\footnotemark,
Y.~Shao$^{(3)}$\footnotemark, 
B.~Srivastava$^{(5)}$,  T.~J.~M.~Symons$^{(4)}$, M.~Tincknell$^{(5)}$,
C.~Tuv\'e$^{(2)}$, S.~Wang$^{(3)}$, P.~Warren$^{(5)}$,  H.~H.~Wieman$^{(4)}$,
T.
Wienold$^{(4)}$, K.~Wolf$^{(6)}$
 }
\address{
$^{(1)}$ University of California, Davis, California 95616\\
$^{(2)}$ Universit\'a di Catania \& INFN-Sezione di Catania, Catania, Italy,
95129\\
$^{(3)}$ Kent State University, Kent, Ohio 44242\\
$^{(4)}$ Nuclear Science Division, Lawrence Berkeley National Laboratory, 
Berkeley, California 94720\\
$^{(5)}$ Purdue University, West Lafayette, Indiana 47907-1396\\
$^{(6)}$ Texas A\&M University, College Station, Texas 77843 \\ 
(The EOS Collaboration)\\
}
\maketitle
\vskip-.45in
\begin{abstract}
We study the  energy dependence of collective (hydrodynamic-like) nuclear 
matter flow in 400-1970 A MeV Ni+Au and 1000-1970 A MeV Ni+Cu reactions.
 The flow increases with energy, reaches a maximum, and then gradually 
decreases at higher energies.  A way of comparing the energy dependence
 of flow values for different projectile-target mass combinations is 
introduced, which demonstrates a common scaling behaviour among flow values 
from differerent systems.
\end{abstract}
\newpage
%
%
\narrowtext

 The study of nuclear matter over a wide range of temperatures and densities
 and the determination of the Equation of State (EOS) of nuclear matter 
continue to be of considerable interest\cite{1.}.  Lacking the possibility 
of comprehensive studies of nuclear matter in bulk (as in neutron stars), 
one resorts to the study of transient, finite systems  provided by 
nucleus-nucleus collisions (over a wide range of energy).  It is now clear
 that the extraction of EOS information from nuclear collisions requires a
 comprehensive set of measurements of collision observables, which can be 
compared to realistic microscopic calculations involving the nuclear matter
 variables.
	
At GeV per nucleon energies, where the collision velocity exceeds that of 
nuclear sound, the collisions produce densities several times higher than 
ground state densities and exhibit compression-induced flow of nuclear 
matter\cite{1.}.  However, it was not until the analysis of events from 
the 4$\pi$ Plastic Ball/Wall\cite{2.} and Streamer Chamber \cite{3.} 
detector systems at the Bevalac that the matter flow characteristics could be 
studied and quantified for a range of systems and energies.  More recently, 
at the Bevalac at LBNL, the EOS Collaboration carried out a comprehensive set
 of measurements over a wide range of energies, and projectile-target 
combinations\cite{4.}.  Similar studies are underway at GSI in Darmstadt and,
 with lighter projectiles, by the DIOGENE Collaboration at Saclay\cite{5.}. 
EOS Collaboration data has been used to study flow for the $Au + Au$  system
 at lab energies ranging from 250 to 1150A~MeV\cite{6.}.  Particle flow for 
protons, deuterons and alpha particles has been determined, using the 
transverse momentum method\cite{7.}. 
The flow is found to increase with particle mass A, and with energy up to
 projectile energies of 1150A~MeV where it tends to level out at values 
close to the Plastic Ball data\cite{8.}. 

Here we present analyses of recent EOS $Ni + Au$ data with energies between
 400 and 1970A~MeV and EOS $Ni +Cu$ data at energies of 1000, 1500 and 
1970A~MeV. This study with the Ni beam allows the use of higher energy 
per nucleon projectiles, and so extends the energy of our flow measurements
 beyond the 1150A~MeV limit of the EOS $Au+Au$ data\cite{6.}.  We also 
present a comparison of the flow values with predictions of a BUU model
 from Bauer~$et~al.$\cite{9.}, and introduce a scaled flow  which allows
 the comparison of flow data from a variety of projectile-target mass systems.
 In total, the data (supported by the calculations) provide the strongest 
evidence yet that, with increasing energy, flow reaches a maximum near 
1000A~MeV, and then declines.

 The EOS Collaboration detector systems have been described  in Refs.~4~and~6.
 The data presented here were obtained using the EOS Time Projection 
Chamber\cite{10.}, situated in the magnetic field of the HISS Magnet.  The 
TPC provides fairly unambiguous particle identification, as well as a 
measurement of momentum, for particles of charge up to $Z=6$.  Particle ID is
 ambiguous for rigidities above $2.4~$GeV/c. 
 Thus, some misidentification will occur and its effect is
 represented in the uncertainties. The target is just upstream from the TPC
 and this results in a large and nearly seamless acceptance.  
Laser beam calibrations provide a check of the corrections for B field 
inhomogeneities, and measurements of drift velocity.   Simulations have been 
performed to study the geometrical acceptance of the detector and to provide
 acceptance corrections.
    
In the present experiment, the thickness of the $Au$ target was 730 
mg/cm$^2$ corresponding to about 1\% interaction probability.  The trigger 
was provided by a scintillator just downstream from the target.  The MUSIC
 (Multiple Sampling Ionization Chamber) detector, downstream from the TPC,
 provided an on-line check on the threshold charge of the heaviest fragment
 allowed by the trigger. For this analysis, data for a wide range of impact
 parameters were taken.  We used charged particle  multiplicity  as a measure
 of the collision centrality, and have adopted the Plastic Ball\cite{8.} 
convention by  dividing the events into five multiplicity bins with bin M1
 corresponding to the most peripheral and bin M5 having the most central 
events.  For the $ Ni + Au $ 400, 600, 1000 and 1970A~MeV the numbers of 
events analyzed were 32k, 26k, 33k and 26k respectively.  The $ Ni + Cu $
 1000, 1500  and 1970 A MeV analyses were done on recently analyzed data 
sets of respectively, 35k, 19k and 48k events. For each event, the reaction
 plane was determined using the transverse momenta of the particles, as 
proposed by Danielewicz and Odyniec\cite{7.}.  The plane is determined by
 the vector $\vec Q=\sum_i w_i\vec p\;_i^t$ and the incident beam direction.
 Here $\vec p\;_i^t$ is the transverse momentum of particle $i$, and $w_i$ 
is a weighting factor defined to maximize the contribution of high rapidity
 particles to the $\vec Q$ vector determination. Normally, for symmetric 
collisions of equal $A$, $w_i$ is taken to increase linearly with rapidity
 up to $w_i=\pm 1$ at $|y'_{\rm cm}|=\delta$ and to be constant at $\pm 1$
 for $|y'_{\rm cm}|\geq\delta$.  Using the sub-event method of Ref.~7 we 
found the optimal reaction plane determination to be made with $\delta\simeq 0.
7$.  We also found that varying $\delta$ by 0.1 results in only a 1\% change 
in the flow values.   For the asymmetric $^{58}Ni\,+\,^{197}Au$, the 
nucleon-nucleon and nucleus-nucleus center of mass frames are different.
  There is therefore some ambiguity as to which reference frame to use for
 determining the weighting factors. We went through an iterative procedure
 as follows:  We start out in the nucleus-nucleus cm frame,
 do the $\vec Q$ weighting in that frame, and then project the $\vec p_t$ of
 each particle onto the reaction plane. We refer to this projection as $p_x$
 for each particle in the event. Plotting the $\langle p_x/A\rangle$ 
(henceforth referred to as $\tilde p_x$) vs. rapidity for all events at a
 given energy then yields the typical S-shaped curves.  These S curves do
 not cross the $\tilde p_x=0$ axis at $y_c=0$  where $y_c$ here refers to
 the rapidity in the chosen  frame, in this case the nucleus-nucleus cm frame.
 We next look at $\tilde p_x$ vs. lab rapidity and note where the curve 
crosses the $\tilde p_x=0$ axis.  This crossing rapidity defines the velocity 
of the new frame in which  we do the next weighting. A couple of iterations 
yields a stable crossing value of $y$.   Finally, we need to correct $\tilde
 p_x$ for the fact that  we project onto an imperfectly known  reaction plane.
  For this we use the sub-event method of Ref.~7.  These  corrections increase
 the $\tilde p_x$ values from 10\% (for the more central $ Ni + Au $ events) 
to 20\% (for the more peripheral $ Ni + Cu $ events).

Figure~1 shows the S-shaped plots of $\tilde p_x$ vs. $y'$  for the four $ 
Ni + Au $ systems and for multiplicity bin M4.   Here, for the moment, we 
include only protons, for which  we have  unambiguous particle identification.
  Following the Plastic Ball analysis  we define the flow, $F$, as the slope 
$(d\tilde p_x/dy')_{\tilde p_x=0}$ at the  zero crossing (generally around 
$y'=0.35$). The slope is calculated from a linear fit to the data. Similar 
$F$ values are obtained using a cubic fit. The imperfect asymmetry of the 
S-curve with respect to the $\tilde p_x=0$ axis is due to both the lower 
acceptance in the lower rapidities, as well as reflecting transverse momentum
 conservation in the asymmetric collision: The larger number of (mainly target)
 particles (both spectator and participant) at low rapidity, is balanced at 
higher rapidity, by fewer particles having larger average $p_t$ values.
 From the geometrical acceptance studies we found that acceptance corrections
 on the flow values are on the order of 5\% for the lower energy $ Ni + Au $ 
400A and 600A~MeV systems . Flow value corrections above 600A~MeV are
 negligible since the detector has full acceptance for the region where
 the fitting is performed.
  
In Figure~2 the extracted values of the slope, $F$ (for protons), at $y'(p_x=0)
$ are plotted vs. projectile kinetic energy per nucleon and compared to the 
flow  predicted by a basic BUU model\cite{9.}. 
 This model does not include composite particle formation so the comparison
 is with the experimental proton $F$ values.  With the inclusion of bound 
protons (from d, t, etc.) the trend of the $F$ values, as a function of energy,
 is similar. It should be noted that the BUU model used here does not include 
momentum dependent interactions.  It is expected that adding these will 
increase the predicted BUU flow values, bringing them to better agreement
 with the data.   The hard EOS BUU flow values appear to reach their maximum 
at a lower energy than the soft EOS. In this aspect the hard EOS BUU is closer
 to the energy dependence of the data. However, neither case gives 
quantitatively the energy dependence, and work on this aspect 
continues\cite{11.}. The main point here is that the BUU model does predict
 a rise and fall of flow. The energy dependence in Fig.~2 is consistent with 
the EOS $Au + Au$ proton flow data of Partlan~$et~al.$\cite{6.} and the 
Plastic Ball data\cite{8.} which show the flow beginning to plateau above
 800A~MeV.

We have studied other flow observables:  the maximum value of $\tilde p_x$,
 the maximum value of $F$, regardless of the corresponding $y$ value, and the 
sum $\sum {\tilde p}_x$, summed over all values for $y'>y'(\tilde p_x=0)$. 
 All of these observables give a flow energy dependence consistent with
 that for $F$.  We have also analyzed the data using the reaction 
plane independent flow signal quantity\cite{12.} proposed by the FOPI group. 
 In addition we
have examined the flow angle dependence on multiplicity for the 4 energies. 
 Both the flow angle and the FOPI signal show an energy dependence (rise and
 fall)
similar to that of $F$ in Fig.~2.  

It can be argued that flow should be determined from the laboratory rapidity,
 $y$, rather than from reduced rapidity, $y'=y/y_{\rm beam}$. Then flow 
becomes  $F_y=d\tilde p_x/dy=F/y_p$ where $y_p$ is the projectile beam 
rapidity in the lab frame.  Plotting $F_y$~vs.~energy we find that the rise 
of flow is reduced while the decline becomes more significant.  
	
The rise and fall of flow is explicable or at least qualitatively reasonable. 
As energy increases, the nucleon-nucleon cross sections become more forward 
peaked. The mean transverse momentum at first rises rapidly with energy and
 then is almost constant above $p_z\simeq 2~$GeV/c.  Thus, flow as a 
manifestation of multiple scattering effects should eventually fall as $p_z$
 continues to increase.
 	
It is of great interest to compare the flow values for a wide range of data. 
 To allow for different projectile-target $(A_1,A_2)$ mass combinations, 
we divide the flow value by $(A_1^{1/3}+A_2^{1/3})$  and call 
$F_S=F/(A_1^{1/3}+A_2^{1/3})$ the scaled flow.  In recent calculations of 
flow\cite{13.}, the authors use an $F/A^{1/3}$ scaling for symmetric systems, 
and argue that, for a given energy, the flow should scale with collision 
(compression) time.  This suggested the $(A_1^{1/3}+A_2^{1/3})^{-1}$ scaling 
used here.  Figure~3 shows a plot of $F_S$ vs. energy per nucleon of the
 projectile.  We include here data from the EOS experiment (solid points), 
along with values derived from other experiments\cite{{8.},{14.},{15.},{16.}} 
for a variety of energies and mass combinations.
   As closely as possible the data selected correspond to Plastic Ball 
multiplicity bins M3 and M4 or to an equivalent range of impact parameters. 
For the EOS  and Plastic Ball data all the isotopes of $Z=1$ and 2  are 
included, except for \cite{16.} where the data is for Z=1 and multiplicity
 bin M3. The Streamer Chamber data\cite{{14.},{15.}} normally include all 
protons, whether free or bound in clusters.   Generally  the flow values 
using all isotopes of $Z = 1$ and 2 are $10-20\%$ larger than those for 
protons. 
In Fig. 3 the scaled flow values, $F_S$, follow, within the uncertainties,
 a common trend with an initial steep rise and then an indication of a gradual
 decline.  The Plastic Ball data are quoted with fairly small statistical 
uncertainties, $\simeq 4-7\%$, about twice the size of the data points in 
Fig. 3. For the $Ar + Pb$\cite{14.} and $Ar + KCl$\cite{15.} Streamer Chamber
  data, we estimated the flow and statistical uncertainties from the $\tilde 
p_x$ vs. $y$ data plots, for the appropriate multiplicity ranges.  Our 
estimated uncertainties for these flow values are in the range of 15-23\%.
 The 1800A~MeV  $Ar + KCl$ Streamer Chamber data, as analyzed in Ref. 7,
 produce a very large flow and scaled flow value ($F_s \simeq 93$~MeV/c 
per nucleon which is off the plot scale and not included in Fig.~3).
 
In summary, we have determined the nucleon flow for Ni induced  collisions
 over an energy range of 400 to 1970A~MeV.  For these Ni systems flow, $F$,
 as measured by the change with rapidity of the average transverse momentum,
 rises with energy, and then declines. Comparison of our flow results with 
flow data from other mass systems is made by introducing a scaled flow, 
$F_S=F/(A_1^{1/3}+A_2^{1/3})$  which exhibits a nearly universal flow 
energy dependence. 

\acknowledgements
	We are pleased to acknowledge the support of the National
Science Foundation (Grant No. PHY-9123301), the Department of Energy
(Contracts/grants DE-AC03-76SF00098, DE-FG02-89ER40531, DE-FG02-88ER40408, 
 DE-FG02-88ER40412, DE-FG05-88ER40437) and Associated
Western Universities; and the assistance of the Bevalac Operations Support
Groups.

\begin{description}
\item[${}^*$] Present address: Sung Kwun Kwan University, Suwon, Rep.
of Korea, 440-746
\item[${}^\dag$] Present address: State University of New York,
Stonybrook, NY 11794
\item[${}^\ddag$] Present address: Lawrence University, Appleton, WI
54912
\item[${}^\S$] Present address: Nuclear Science Division, LBNL, Berkeley, CA
94720
\item[${}^{**}$] Present address: The Svedberg Laboratory, University
of Uppsala, S751-21 Sweden
\item[${}^{\dag\dag}$] Present address: Physics Department, College of
Wooster,
Wooster, OH 44691 
\item[${}^{\ddag\ddag}$] Present address: Crump Institute for Biological
Imaging,
UCLA, Los Angeles, CA 91776
\end{description}

\begin{figure}
\psfig{file=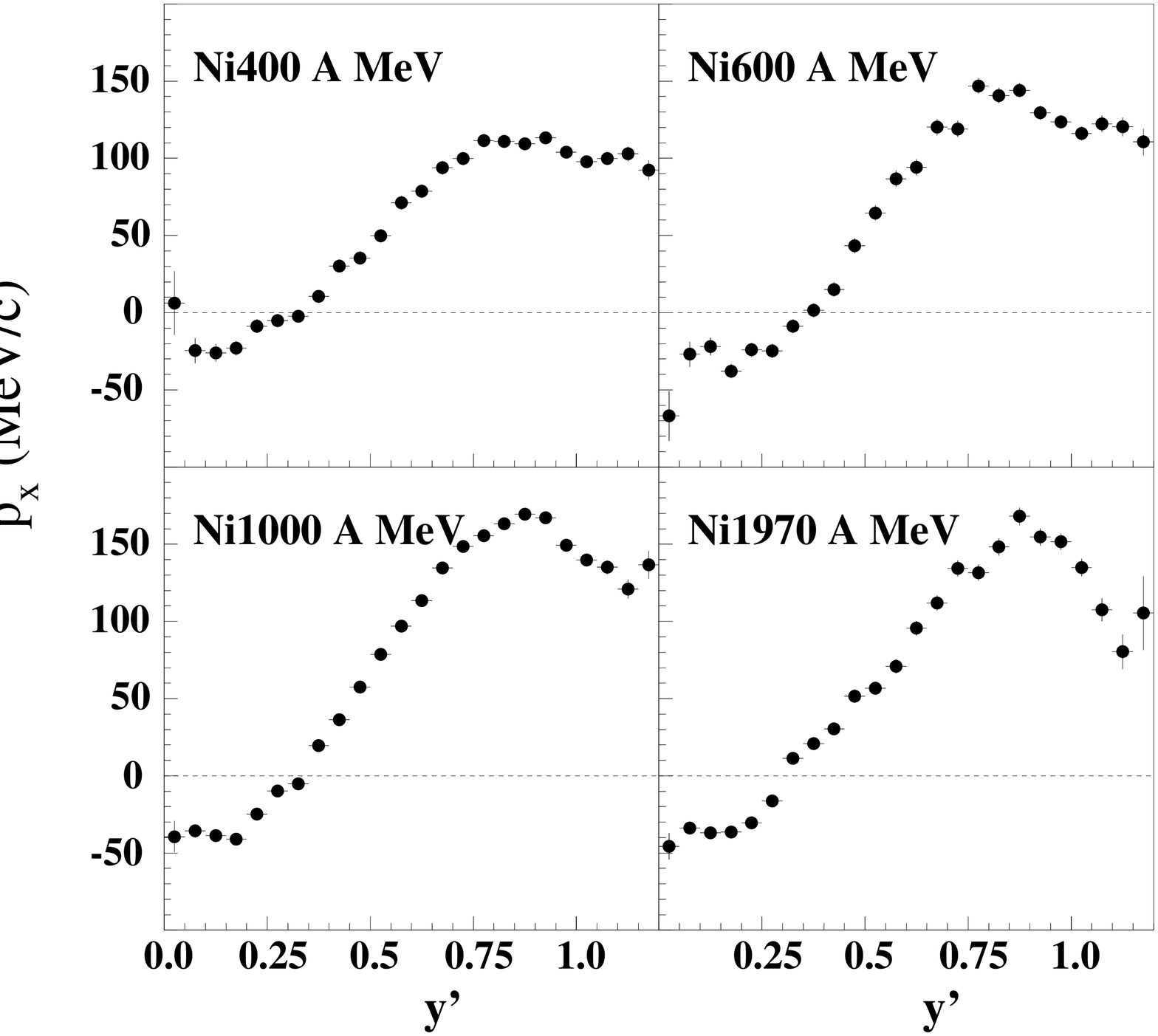,width=3in,height=3in}
\caption{S curves for the four $Ni+Au$ energies, using protons only in
multiplicity bin M4.
	  $\tilde p_x$  is the average of the
{\it x} component of the momentum (see text). $y'$ is the
rapidity in the lab frame scaled by the
 rapidity of the beam.} 
 \label{FIG. 1}
\end{figure}

\begin{figure}
\psfig{file=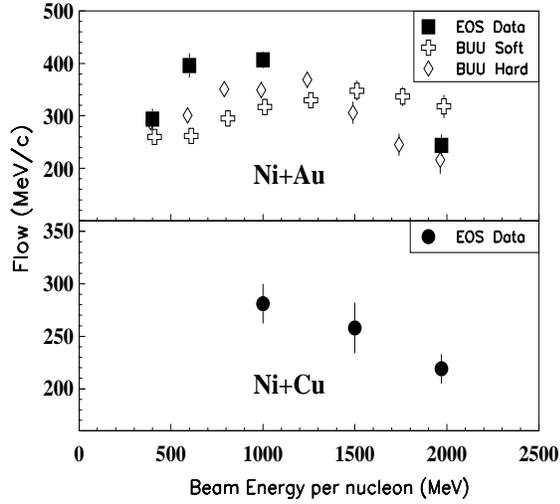,width=3in,height=3in}
\caption{	Proton flow as a function of beam energy per nucleon in 
multiplicity bin M4. 
The top panel is for $Ni+Au$ EOS data. The solid
symbols are the flow values from the graphs in Fig.~1 while the open symbols
 are for soft and hard equation of state BUU $Ni+Au$ calculations with an 
equivalent impact parameter distribution.  The bottom panel consists of 
flow values determined from $Ni+Cu$ EOS data.}
\label{FIG. 2}
\end{figure}

\begin{figure}
\psfig{file=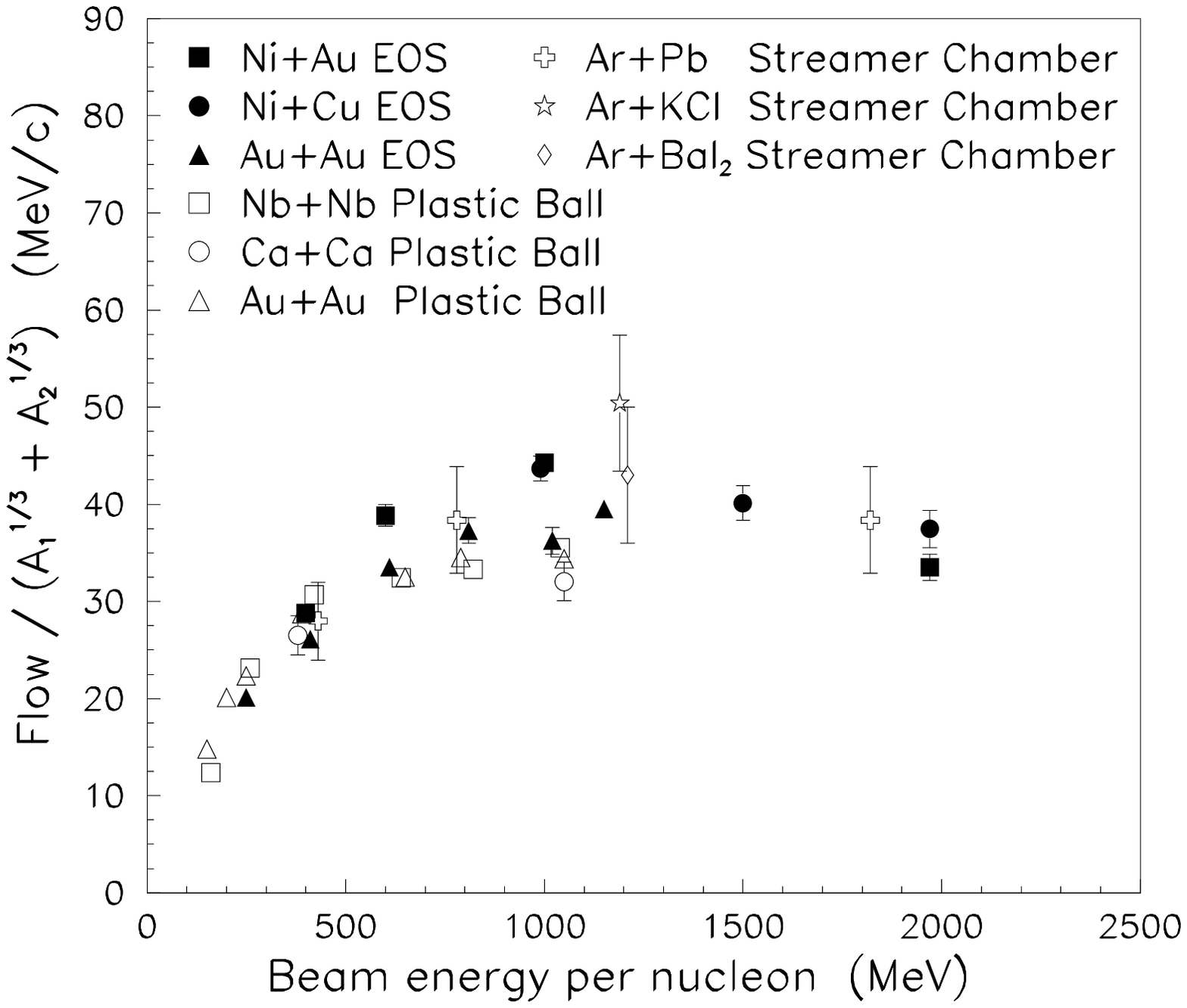,width=3in,height=3in}
\caption{	Scaled flow values  vs. beam energy per
nucleon  for different projectile-target systems for Plastic Ball multiplicity
 bins 3+4. In the EOS and Plastic Ball data all isotopes of Z=1,2 are included.
  For the Streamer Chamber all free and bound protons are included.  
To improve the distinction between data points at the same beam energy, some
 of the beam energy values have been staggered around by as much as 20 MeV.}
\label{FIG. 3}
\end{figure}


\begin{references}

\bibitem{1.}	A recent conference on this topic has been held.  Please see 
the Nuclear Equation of State, Vol. 216 of NATO Advanced Study Institute, 
Series B: Physics, edited by W. Greiner and H. St\"ocker (Plenum, New York,
 1989).

\bibitem{2.}	H.~\AA. Gustafsson {\it et al.}, Phys. Rev. Lett. {\bf 52},
 1590 (1984).

\bibitem{3.}	R.~E. Renfordt {\it et al.}, Phys. Rev. Lett. {\bf 53},
 763 (1984); H. Str\"obele {\it et al.}, Phys. Rev. C {\bf 27}, 1349 (1983).

\bibitem{4.}	S. Costa (for the EOS Collaboration), Proceedings of the 23rd
 International Winter Meeting on Nuclear Physics at Bormio, Italy, January 
23-30 (1994).

\bibitem{5.} J. Gosset {\it et al.}, ibid. Ref.~1, p.~87; J. Poitou {\it 
et al.}, Nucl. Phys. A {\bf 536}, 767 (1992).

\bibitem{6.}	M. Partlan {\it et al.} (the EOS Collaboration), Phys. Rev. 
Lett. {\bf 75}, 2100 (1995). 

\bibitem{7.}    P. Danielewicz and G. Odyniec, Phys. Lett. {\bf 157B}, 146 
(1985).

\bibitem{8.}	H.~H. Gutbrod, A.~M. Poskanzer and H.-G. Ritter, Reports on
 Progress in Physics {\bf 52}, 1267 (1989), and references therein. 

\bibitem{9.}	W. Bauer, Phys. Rev. Lett. {\bf 61}, 2534 (1988); Bao-An Li
 and W. Bauer, Phys. Rev. C {\bf 44}, 2095 (1991); W. Bauer, C.~K. Gelbke, 
and S. Pratt,  Ann.  Rev.  Nucl.  Sci.  {\bf 42}, 77 (1992).

\bibitem{10.} G. Rai {\it et al.}, IEEE Trans. Nucl. Sci. {\bf 37}, 56 (1990).

\bibitem{11.} J.~Chance, Ph.D. Dissertation, UC Davis, 1996.

\bibitem{12.} T. Wienold {\it et al.}, GSI-Preprint-94-74.

\bibitem{13.}  A. Lang, B. Bl\"attel, W. Cassing, V. Koch, U. Mosel and K.
Weber, Zeitschrift f\"ur Physik A {\bf 340}, 287 (1991).

\bibitem{14.}	D. Beavis {\it et al.}, Phys. Rev. C {\bf 45}, 299 (1992).

\bibitem{15.} D. Beavis, S.~Y. Chu, S.~Y. Fung, W. Gorn, D. Keane, Y.~M. Liu,
G.Van Dalen, M. Vient, Phys. Rev. C {\bf 33}, 1113, (1986).

\bibitem{16.} J.W. Harris et al., Nucl. Phys. A {\bf 471}, 241C (1987); K.G.R. 
Doss et al., Phys. Rev. Lett. {\bf 59}, 2720 (1987).

\end{references}
\end{document}